\documentclass[arxiv,usenatbib]{iupartex}

\usepackage{newtxtext,newtxmath}
\usepackage[T1]{fontenc}
\usepackage{float}
\usepackage{adjustbox}
\usepackage{amsmath, amssymb, amsfonts, bm}
\usepackage{mathrsfs}
\usepackage{orcidlink}
\usepackage{subcaption}

\title[]{Complex-Potential Analysis of a Slit Model in Electrowetting}

\author[Shah et al.]{%
Dhairya Shah$^{1}{\orcidlink{0000-0002-4496-7737}}$, %
Yuan Liu$^{1,2}{\orcidlink{0009-0008-3208-8306}}$, 
Samuel Brzezicki$^{1\cc}$
\affsep \\
$^1$Department of Mathematics, Imperial College London, 180 Queen’s Gate, London
SW7 2AZ, United Kingdom\\
$^2$Journal of Shandong University of Finance and Economics, No 7366, Erhuandong Road, Jinan 250014, China
}

\corres{Samuel Brzezicki}{samuel.brzezicki10@imperial.ac.uk}

\begin{document}
\label{firstpage}
\pagerange{\pageref*{firstpage}--\pageref*{lastpage}}
\maketitle

\begin{abstract}
This study develops a complex-potential framework for the electric field near the Triple Contact Point in electrowetting. A global slit model provides the electrostatic potential, surface-charge structure and endpoint expansion. The resulting solution also gives an analytical approximation to the droplet shape and exhibits the characteristic edge singularity of the zero-contact-angle slit model.
\end{abstract}



\section{Introduction}
The contact angle between the liquid-vapour and solid-liquid interfaces, referred to as 
Young’s angle \cite{young_1805}, measures the wettability of solid surfaces. Altering the hydrophilicity via an external voltage is known as electrowetting, a process in which the wettability of a solid with respect to a liquid is enhanced by generating a voltage difference between the liquid and the solid substrate. Due to the electric stress applied by the external electric field, both the contact angle and the curvature of the droplet's surface alter \cite{Taylor_64}. This phenomenon has gained contemporary attention because of its relevance in microfluidics applications, including everyday uses like hydrophobic coatings \cite{Sushanta_18}, liquid shutters \cite{Lee_21}, and engineering devices such as droplet-based energy generators \cite{Wu_20} and electrowetting displays \cite{Yong_17}. However, the theoretical understanding of the Triple Contact Point (TCP) remains incomplete. In his pioneering work on electrowetting, Lord \citet{Rayleigh_1882} found that adding charge to a droplet beyond a critical value can cause it to split, which relates to TCP singularity. Macroscopic models, such as the Young-Lippmann equation, notoriously break down near the TCP due to the onset of a mathematical singularity in the electric field and stress. 

The behavior of electrical droplets' boundary curves and surfaces has been researched through experimental and numerical studies, physical approaches, and theoretical analyses. Experiments by \cite{quin_05} revealed that contact angle saturation in electrowetting—the droplet surface reaching a limiting angle under stronger external voltage—is determined by the material properties of the system. \citet{MugeleF2007} presents evidence that, on a small scale, the contact angle remains fixed even when subjected to a varying voltage. \citet{BerozJ_19} demonstrated experimentally and theoretically a power law governing the stability limit of a conducting droplet or bubble exposed to an external electric field.

This inherent complexity of boundary problems motivates the use of advanced analytical tools. \citet{crowdy2015} proposed a conformal mapping for the $90^\circ$ contact angle droplet, building upon \cite{Crowdy1999,Crowdy2000}. Inspired by the approach in \cite{10.1063/1.4821137} and the power of complex analysis in simplifying and solving electrowetting problems, we construct a slit-based complex-potential model for the electrowetting problem and examine its electrostatic field and surface-charge structure.

Motivated by these considerations, we construct a global slit model using complex analysis. The complex potentials in the air and dielectric regions provide the electrostatic field, surface-charge distribution and endpoint structure, and are further used to obtain an analytical approximation to the droplet shape.
\section{Slit Model and Complex Potential}
\subsection{Far-field Complex Potential}
\begin{figure}
        \centering
        \includegraphics[width=.8\linewidth]{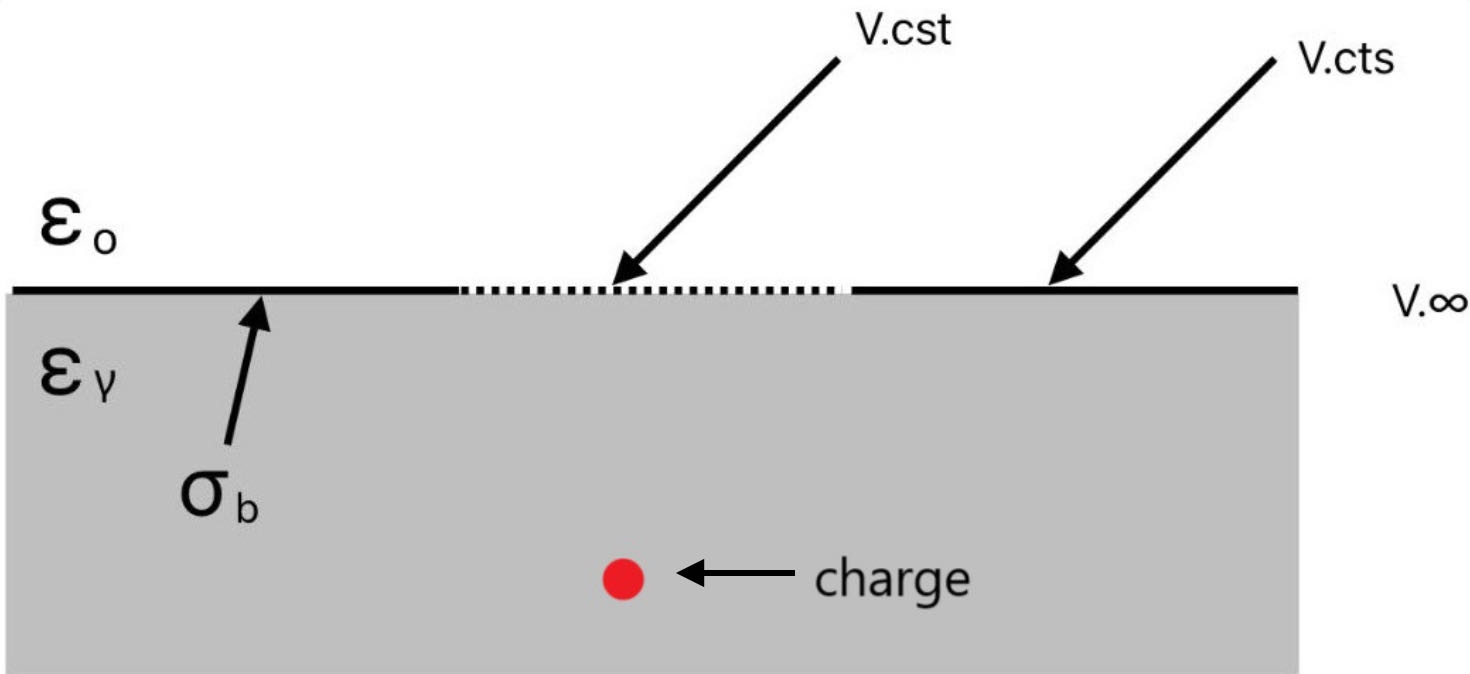}
        \caption{
        The model depicts a thin droplet slit (dashed line) on a dielectric layer (shaded area, permittivity $\epsilon_{\gamma}$) with air above (blank space, permittivity $\epsilon_0$). The slit itself is an equipotential surface (V.cst). The voltage is continuous (V.cts) along the dielectric-air interface (thick black lines). An induced bound surface charge density $\sigma_b$ resides on the dielectric surface. The electrode effect is simplified to a variable point charge (red spot) embedded within the dielectric. V.$\infty$ is the far-field voltage.}
        \label{fig:slit}
    \end{figure}

Consider a droplet situated on a dielectric layer, with electrodes altering the droplet shape and contact angle. When the droplet is observed from a large distance, its shape simplifies into a thin film of negligible height, which we refer to as a "slit", where the contact angle at large distances approaches $0^\circ$. The dielectric layer is assumed to be semi-infinite in the $-y$ and $\pm x$ directions. The electrodes are modeled as a point charge. Figure \ref{fig:slit} illustrates this simplification.

We denote the dielectric permittivity by $\epsilon_\gamma$ and its relative permittivity by $\epsilon_r=\epsilon_\gamma/\epsilon_0$.
    
The excess charges of the conductive droplet $\Omega$ form a surface charge density on its boundary $\partial\Omega$. The voltage must be continuous across all domain interfaces, specifically the droplet interfaces $\partial \Omega$ and the dielectric interfaces $\partial\mathcal{D}$ 
    \begin{equation}\label{eqn:v.cts}
\left.V_{above}\right|_{\partial\mathcal{D}^+}=\left.V_{below}\right|_{\partial\mathcal{D}^-}     
    \end{equation}
 $V_{above}$ is the electric potential in the air above, and $V_{below}$ is the potential in the dielectric $\mathcal{D}$. No free charge is placed on the dielectric--air interface, so the normal electric displacement is continuous. Using the same $+y$ direction on both sides,
 \begin{equation}\label{eqn:gauss}
    \epsilon_0 E_{y,above}=\epsilon_\gamma E_{y,below}.
    \end{equation}
The induced bound polarisation charge on the dielectric surface is
\[
\sigma_b=\epsilon_0(E_{y,above}-E_{y,below})=(\epsilon_\gamma-\epsilon_0)E_{y,below}.
\]
Here, $\vec{E}_{above}$ and $\vec{E}_{below}$ represent the electric fields in the air and the dielectric, respectively, and both normal components use the same $+y$ direction.

To simplify the determination of the electric potential, we employ the Joukowski transformation. The original physical domain containing the slit is mapped to a unit disc in the $\zeta$-plane. Let $w$, $w_{above}$ and $w_{below}$ denote the complex potentials where $V=\Im [w]$. The boundary conditions include the following:
\begin{itemize}
    \item $\nabla^2 w = 0$ everywhere, except for the location of the point charge.  
    \item In the conducting droplet $|\zeta|\leq1$, the voltage is constant: $\Im [w]=0$.
    \item At the dielectric interface $\zeta\in\mathbb{R}$, the voltage is continuous: $\Im [w_{above}]=\Im [w_{below}]$.
    \item At the dielectric interface, the normal electric displacement is continuous.
\end{itemize}
To ensure the unit circle is an equipotential in the presence of a point charge at $\zeta_0$, the conductor droplet will induce an image potential
\begin{equation}
\overline{w_0}(\frac{1}{\zeta})=-\mathrm{i} \log (\frac{1}{\zeta}-\bar{\zeta_0})
\end{equation}

\begin{figure}[H]
        \centering
        \includegraphics[width=1.\linewidth]{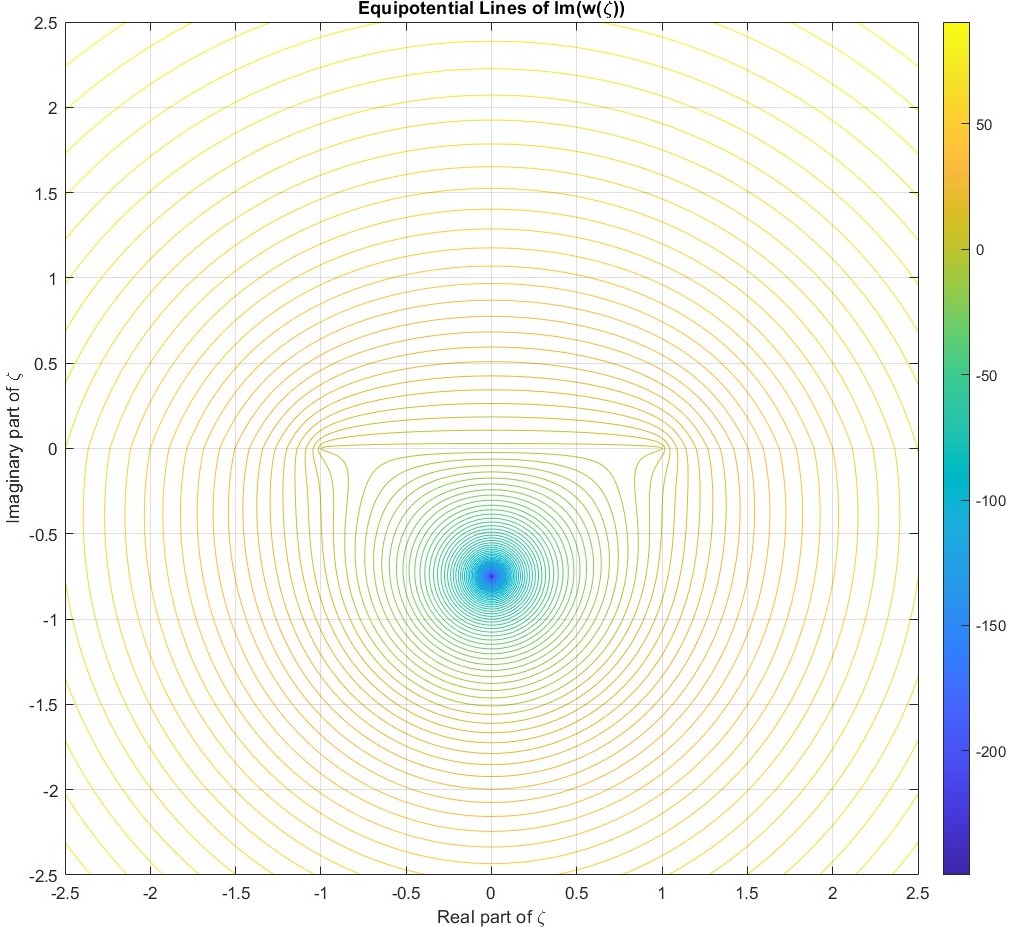}
        \caption{
        The equipotential contours associated with the global complex potentials $w_+(z)$ and $w_-(z)$. The source charge belongs to the dielectric-side construction. Dense contours occur near the two slit endpoints. A slight voltage distortion is visible along the air--dielectric interface due to the induced bound surface charge density.
        }
        \label{fig:global complex potential}
\end{figure}

Hence, the complex potential in ($\text{Im}[\zeta]>0\cap |\zeta|\geq1$) is the sum of these charges
\begin{equation}\label{eqn:w_up}
    w_{+}(\zeta) = \frac{q}{\pi \epsilon_0}\frac{1}{\epsilon_r+1} \mathrm{i} \left[ \log(\zeta - \zeta_0) - \log\left(\frac{1}{\zeta} - \overline{\zeta_0}\right) \right]
\end{equation}
We utilize the Joukowski transformation $\zeta=z\pm\sqrt{z^2-1}$ and select the appropriate branch cut. This mapping transforms the problem from the $\zeta$-plane back to the physical $z$-plane. Consequently, the complex potential in the air region of the physical $z$-plane ($y>0$) is then given by
\begin{equation}\label{eqn:w+z}
w_{+}(z) =\frac{q}{\pi\epsilon_0}\frac{1}{\epsilon_r+1}  \mathrm{i} \left[\log\left(z+\sqrt{z^2-1} - \zeta_0\right) - \log\left(\frac{1}{z+\sqrt{z^2-1}} - \overline{\zeta_0}\right)\right]
\end{equation}
We derived the complex potential in ($\text{Im}[\zeta]\leq 0 \cap |\zeta|\geq 1$) by introducing image charges proportional to the original source terms. These image charges must satisfy the boundary conditions and are scaled by a factor related to the permittivity ratio $\epsilon_r$.
\begin{equation}\label{eqn:w_down}
    w_{-}(\zeta) = \frac{q}{2\pi \epsilon_0\epsilon_r} \mathrm{i} \left[ \log(\zeta - \zeta_0) - \log\left(\frac{1}{\zeta} - \overline{\zeta_0}\right) \right]
+\frac{q}{2\pi \epsilon_0\epsilon_r}\frac{\epsilon_r-1}{\epsilon_r+1}\mathrm{i} \left[ \log(\zeta - \overline{\zeta_0}) - \log\left(\frac{1}{\zeta} - \zeta_0\right) \right]    
\end{equation}
In the dielectric region of the physical $z$-plane region ($y\leq0$), the complex potential $w_{-}(z)$ is hence
\begin{equation}  
\begin{aligned}
w_{-}(z) &= \frac{q}{2\pi \epsilon_0\epsilon_r} \mathrm{i} \left[ \log(z-\sqrt{z^2-1} - \zeta_0) - \log\left(\frac{1}{z-\sqrt{z^2-1}} - \overline{\zeta_0}\right) \right]\\
&+\frac{q}{2\pi \epsilon_0\epsilon_r}\frac{\epsilon_r-1}{\epsilon_r+1}\mathrm{i} \left[ \log(z-\sqrt{z^2-1} - \overline{\zeta_0}) - \log\left(\frac{1}{z-\sqrt{z^2-1}} - \zeta_0\right) \right]
\end{aligned}
\end{equation}
Combining these two parts yields the complete complex potential $w(z)$ across the entire $z$-plane, from which the equipotential lines in Figure \ref{fig:global complex potential} are determined.
\begin{equation}   
\begin{aligned}
w(z)&= \begin{cases}w_+(z) &, \,y>0\\
w_-(z) &,  \,y\leq 0\end{cases}
\end{aligned}\label{eqn:cp_global}
\end{equation}

\subsection{Analytical Droplet Shape}

Following \citet{Fontelos2008_2}, we re-derive the relationship between a droplet's height and its curvature in $x, y$ coordinates, incorporating the Maxwell stress tensor $\mathbf{M}^e$. The total stress tensor, $\mathbf{T}$, satisfies
\begin{equation}
\mathbf{T}=-\Delta p \mathbf{I}+\mathbf{M}^e+\tau
\end{equation}
The analysis of the stress tensor $\mathbf{T}$ yields the following equation (\ref{eqn:height}). Here, $h$ is the droplet height, $\sigma$ is the surface charge density, $\Delta p$ is the pressure difference, and $\tau$ is the surface tension coefficient. 
\begin{equation}
h_{xx}=\Delta p -\frac{q^2}{2}\sigma^2\label{eqn:height}
\end{equation}
The surface charge density $\sigma$ on the upper $z$-plane is computed via the conformal mapping $z=\frac{1}{2}(\zeta+\frac{1}{\zeta})$, where the droplet's upper boundary corresponds to the unit circle $\zeta=e^{\mathrm{i}\varphi}$ in the $\zeta$-plane. The transformation of the surface charge density from the $\zeta$-plane to the $z$-plane is obtained as
\begin{equation}
\sigma(z) = \left| \frac{dw}{d\zeta} \frac{d\zeta}{dz} \right| = \left| \frac{dw}{d\zeta} \right|\frac{1}{\sin \varphi}
\end{equation}
The resulting liquid-air interface shape $h(x)$, applicable for $x\in[0,1]$, is found by solving equation (\ref{eqn:height}) subject to the boundary conditions $h'(0)=0$, $h(1)=0$ and the physical constraint $h\geq 0$. For the specific case of a constant charge density  on the $\zeta$-plane, the analytical droplet shape is given by
\begin{equation}
   h=\frac{\Delta p}{2}x^2+\frac{-\Delta p+q^2 \log2}{2}-\frac{q^2}{2}\left(\frac{1-x}{2}\log(1-x)+\frac{1+x}{2}\log(1+x)\right)
\end{equation}
The observed variation of droplet height with increasing charge $q$ is consistent with the research findings reported in \citet{BerozJ_19} and the results illustrated in \citet{Fontelos2008} (specifically, Figure 7).

Considering the varying surface charge density derived from the complex potential $w_+$ from Equation (\ref{eqn:w+z}) and setting the point charge location as $\zeta_0=-a\mathrm{i}$, we obtain the following non-dimensional equation for the surface charge density $\sigma^*$ on the upper slit boundary:
\begin{equation}  
\sigma^*\sim \frac{2\left(1+a\sqrt{1-x^2}\right)}{\sqrt{1-x^2}\left(1+a^2+2a\sqrt{1-x^2}\right)}
\end{equation}
The calculated surface charge distribution is similar to the results reported in \citet{crowdy2015}, as well as the distributions illustrated in Figure 10 of \citet{Fontelos2008} and Figure 2(b) of \citet{Leong2014}.

\begin{figure}
    \centering
    \includegraphics[width=.9\linewidth]{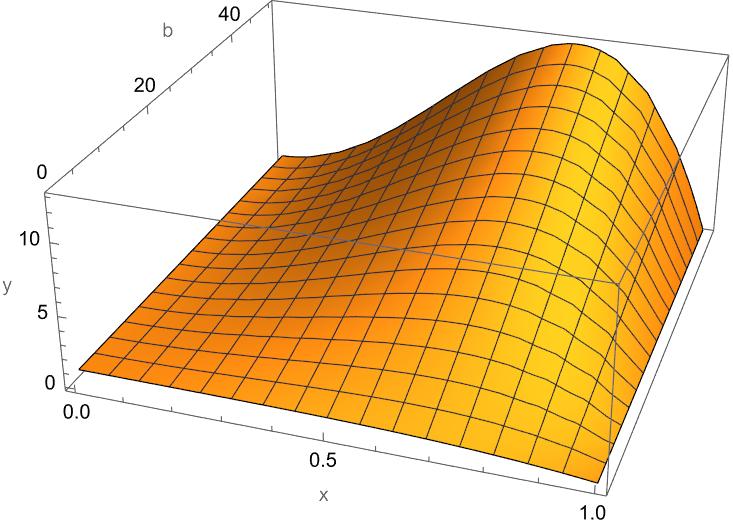}
    \caption{The estimated droplet boundary curve from Equation~(\ref{eqn:h}) for varying values of the shape parameter $B$ (labelled $b$ in the plot).}
    \label{fig:droplet_shape_sigma_vary}
\end{figure}

Truncating the Taylor expansion of $\sigma^2$, we obtain the following approximation
\begin{equation}\label{eqn:h}
    h\approx-(B+1)x^4+B x^2+1   
\end{equation}
The resulting equation is governed by the single parameter $B$. As $B$ increases, the droplet boundary curve shows a corresponding increase in peak height, as depicted in Figure \ref{fig:droplet_shape_sigma_vary}. This observed trend is consistent with the results shown in Figure 2(a) of \citet{Leong2014} and Figure 2.1 of \citet{Fontelos2008_2}.

However, the ideal slit description develops an endpoint singularity in the surface-charge density at $\varphi=0$ and $\varphi=\pi$. This is the edge singularity of the zero-contact-angle slit model and signals that a separate finite-angle local description is required at the TCP.

\subsection{Endpoint Expansion of the Slit Potential}

The macroscale slit model has zero contact angle and does not determine the finite-angle TCP geometry. We use the global solution here to identify its endpoint structure.

\begin{figure}[H]
    \centering
    \begin{tabular}{cc}
    \includegraphics[width=.47\linewidth]{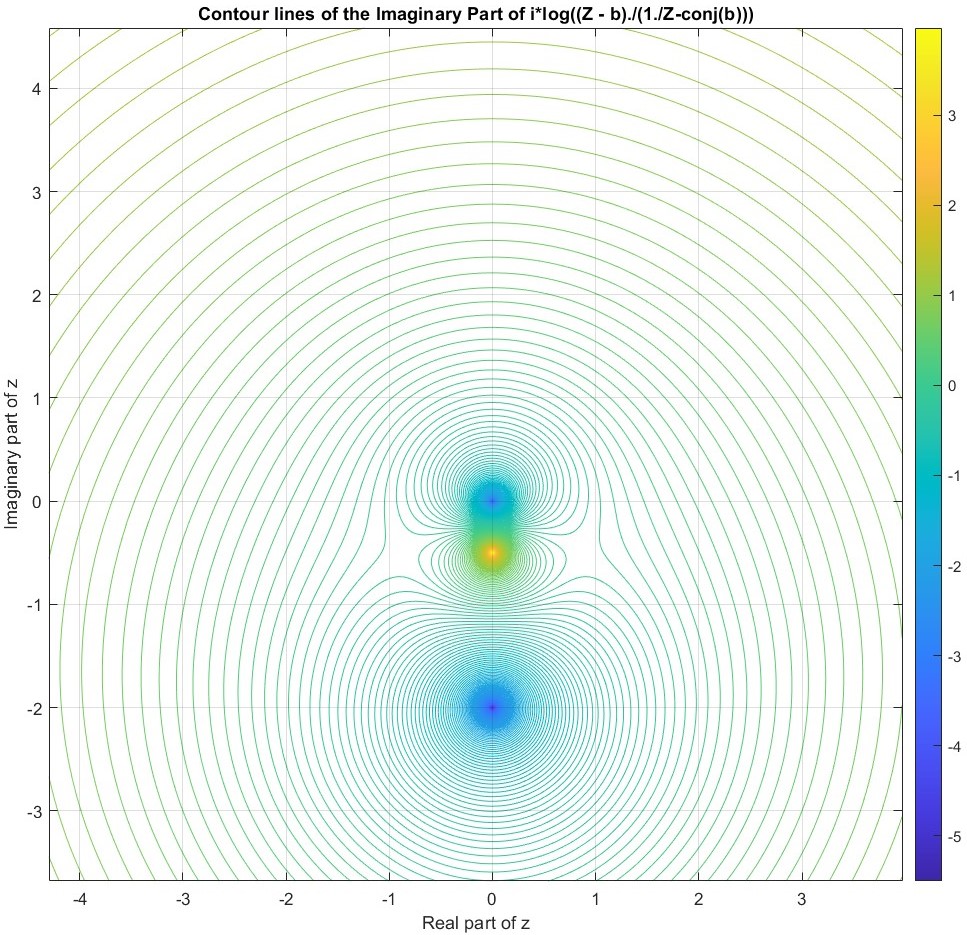}
   &\includegraphics[width=.49\linewidth]{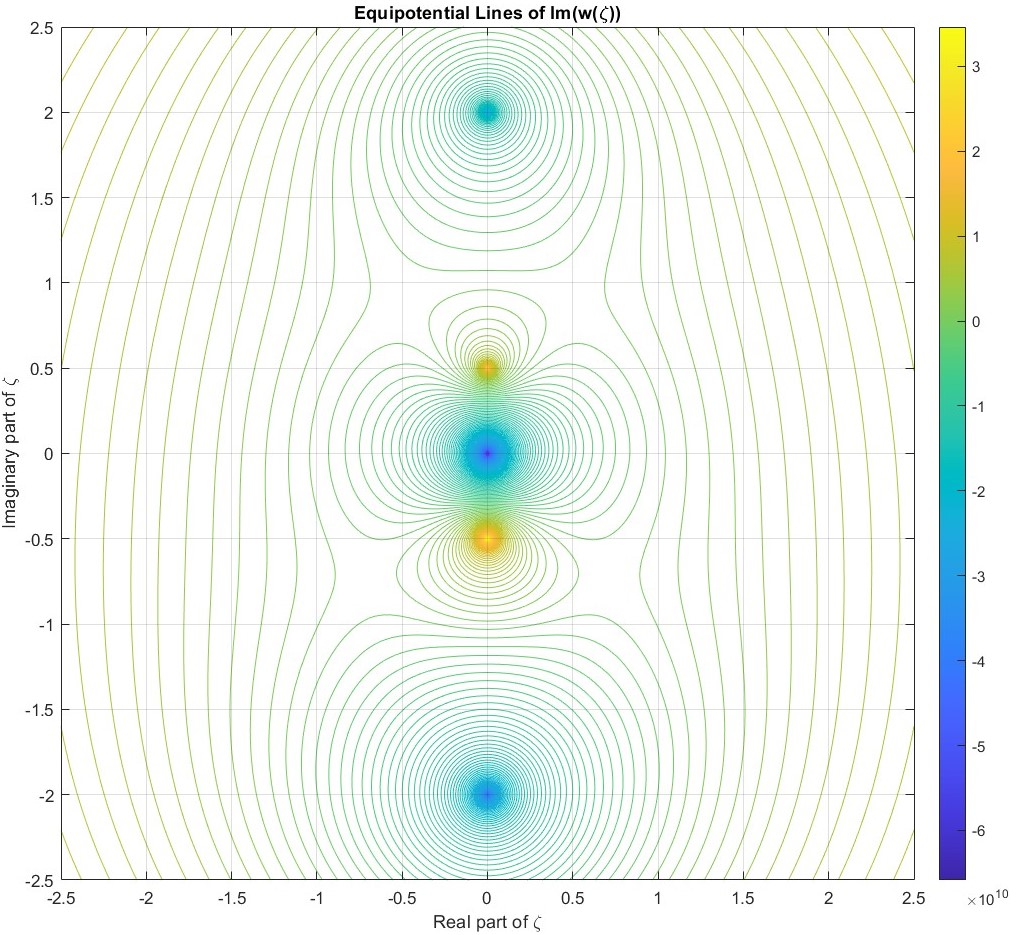}\\
   (a) the equipotential lines in ($\text{Im}[\zeta]>0 \cap |\zeta|\geq 1$)  &(b) the equipotential lines in ($\text{Im}[\zeta]\leq 0 \cap |\zeta|\geq 1$)
   \end{tabular}
    \caption{
    The panels show the plotted continuations associated with the upper and lower complex-potential expressions. The physical air-side field uses the upper expression in $\operatorname{Im}\zeta>0$ with $|\zeta|\geq1$, while the physical dielectric-side field uses the lower expression in $\operatorname{Im}\zeta\leq0$ with $|\zeta|\geq1$.
    }\label{fig:zete_pot}
\end{figure}

In a source-free neighbourhood of the TCP, the influence of the global configuration enters through the coefficients of the endpoint expansion below. The point charge and its images remain part of the global construction; no local point source is introduced.

The asymptotic analysis of the global complex potentials $w_{+}$ and $w_{-}$ near the right-hand TCP, using the local displacement $\xi=z-1$, yields a characteristic series expansion:
\begin{equation}\label{eqn:endpoint_expansion}
w_{\pm}(1+\xi) \approx c_0^{\pm} + c_{1/2}^{\pm} \xi^{1/2} + c_1^{\pm} \xi + c_{3/2}^{\pm} \xi^{3/2} + c_2^{\pm} \xi^2 + \dots
\end{equation}
The coefficients $c_n^{\pm}$ are explicitly determined through the Taylor series expansion. This process reveals that the detailed structure of the external charge source dictates the specific multipole moments permitted in the TCP region.

\section{Conclusion}
We constructed a complex-potential description of the global slit model, obtaining the electrostatic potentials in the air and dielectric regions together with the associated surface-charge structure. The resulting field was further used to obtain an analytical approximation to the droplet shape. The ideal zero-contact-angle slit develops an endpoint singularity, and its local expansion contains the characteristic square-root structure near the slit endpoint.




\bibliographystyle{apalike}
\bibliography{References}

@article{Crowdy2015,
issn = {1070-6631},
journal = {Physics of fluids (1994)},
language = {eng},
author = {Crowdy, Darren},
address = {Melville},
number = {6},
publisher = {American Institute of Physics},
title = {Exact solutions for the static dewetting of two-dimensional charged conducting droplets on a substrate},
volume = {27},
year = {2015},
}

@article{Fontelos2008_2,
author = {Fontelos, M. A. and Kindelán, U.},
copyright = {Copyright 2009 Society for Industrial and Applied Mathematics},
issn = {0036-1399},
journal = {SIAM journal on applied mathematics},
language = {eng},
number = {1},
pages = {126-148},
publisher = {Society for Industrial and Applied Mathematics},
title = {The Shape of Charged Drops over a Solid Surface and Symmetry-Breaking Instabilities},
volume = {69},
year = {2008},
}

@article{Fontelos2008,
copyright = {American Institute of Physics},
issn = {1070-6631},
journal = {Physics of fluids (1994)},
language = {eng},
author = {Fontelos, M. A. and Kindelán, U. and Vantzos, O.},
address = {Melville, NY},
number = {9},
pages = {092110-092110-12},
publisher = {American Institute of Physics},
title = {Evolution of neutral and charged droplets in an electric field},
volume = {20},
year = {2008},
}

@article{Crowdy2000,
copyright = {Massachusetts Institute of Technology 2000},
issn = {0022-2526},
journal = {Studies in applied mathematics (Cambridge)},
language = {eng},
author = {Crowdy, Darren},
address = {Boston, USA and Oxford, UK},
number = {1},
pages = {35-58},
publisher = {Blackwell Publishers Inc},
title = {A New Approach to Free Surface Euler Flows with Capillarity},
volume = {105},
year = {2000},
}

@article{Crowdy1999,
copyright = {American Institute of Physics},
issn = {1070-6631},
journal = {Physics of fluids (1994)},
language = {eng},
author = {Crowdy, Darren},
number = {10},
pages = {2836-2845},
title = {Circulation-induced shape deformations of drops and bubbles: Exact two-dimensional models},
volume = {11},
year = {1999},
}

@article{Leong2014,
copyright = {2014 Author(s). All article content, except where otherwise noted, is licensed under a Creative Commons Attribution 3.0 Unported License.},
issn = {1070-6631},
journal = {Physics of fluids (1994)},
language = {eng},
author = {Leong, Fong Yew and Mirsaidov, Utkur M. and Matsudaira, Paul and Mahadevan, L.},
address = {Melville},
number = {1},
publisher = {American Institute of Physics},
title = {Dynamics of a nanodroplet under a transmission electron microscope},
volume = {26},
year = {2014},
}

@article{MugeleF2007,
author = {Mugele, F and Buehrle, J},
issn = {0953-8984},
journal = {Journal of physics. Condensed matter},
language = {eng},
number = {37},
pages = {375112-375112 (20)},
publisher = {IOP Publishing},
title = {Equilibrium drop surface profiles in electric fields},
volume = {19},
year = {2007},
}

@article{Wu_20,
number = {7},
pages = {},
publisher = {American Physical Society},
title = {Energy Harvesting from Drops Impacting onto Charged Surfaces},
volume = {125},
year = {2020},
language = {eng},
author = {Wu, Hao and Mendel, Niels and den Ende, Dirk van and Zhou, Guofu and Mugele, Frieder},
address = {College Park},
copyright = {Copyright American Physical Society Aug 14, 2020},
issn = {0031-9007},
journal = {Physical review letters},
}

@article{Lee_21,
number = {5},
pages = {055009-055009},
publisher = {American Institute of Physics},
title = {Switchable liquid shutter operated by electrowetting for security of mobile electronics},
volume = {92},
year = {2021},
language = {eng},
author = {Lee, Jeongmin and Park, Yuna and Jang, Deasung and Chung, Sang Kug},
address = {Melville},
copyright = {Author(s)},
issn = {0034-6748},
journal = {Review of scientific instruments},
}

@article{Yong_17,
pages = {754-758},
publisher = {Elsevier B.V},
title = {Smart self-cleaning lens cover for miniature cameras of automobiles},
volume = {239},
year = {2017},
language = {eng},
author = {Yong Lee, Kang and Hong, Jiwoo and Chung, Sang Kug},
address = {Lausanne},
copyright = {2016 Elsevier B.V.},
issn = {0925-4005},
journal = {Sensors and actuators. B, Chemical},
}

@article{BerozJ_19,
pages = {244501-244501},
language = {eng},
number = {24},
publisher = {American Physical Society},
title = {Stability Limit of Electrified Droplets},
volume = {122},
year = {2019},
issn = {0031-9007},
journal = {Physical review letters},
author = {Beroz, J and Hart, A J and Bush, J W M},
address = {United States},
}

@article{Rayleigh_1882,
author = {Lord Rayleigh},
title = {XX. On the equilibrium of liquid conducting masses charged with electricity },
journal = {The London, Edinburgh, and Dublin Philosophical Magazine and Journal of Science},
volume = {14},
number = {87},
pages = {184--186},
year = {1882},
publisher = {Taylor \& Francis},
doi = {10.1080/14786448208628425},
URL = { https://doi.org/10.1080/14786448208628425
},
eprint = {https://doi.org/10.1080/14786448208628425   
}
}

@article{Young_1805,
pages = {65-87},
publisher = {The Royal Society},
title = {III. An essay on the cohesion of fluids},
volume = {95},
year = {1805},
language = {eng},
author = {Young, Thomas},
address = {London},
copyright = {Scanned images copyright © 2017, Royal Society},
issn = {0261-0523},
journal = {Philosophical transactions of the Royal Society of London},
}

@article{Sushanta_18,
author = {Sushanta Kumar Sethi and Gaurav Manik},
title = {Recent Progress in Super Hydrophobic/Hydrophilic Self-Cleaning Surfaces for Various Industrial Applications: A Review},
journal = {Polymer-Plastics Technology and Engineering},
volume = {57},
number = {18},
pages = {1932--1952},
year = {2018},
publisher = {Taylor \& Francis},
doi = {10.1080/03602559.2018.1447128},
URL = {    
        https://doi.org/10.1080/03602559.2018.1447128   
},
eprint = {     
        https://doi.org/10.1080/03602559.2018.1447128 
}

}

@article{Quin_05,
number = {13},
pages = {6268-6275},
publisher = {American Chemical Society},
title = {Contact Angle Saturation in Electrowetting},
volume = {109},
year = {2005},
language = {eng},
author = {Quinn, Anthony and Sedev, Rossen and Ralston, John},
address = {United States},
copyright = {Copyright © 2005 American Chemical Society},
issn = {1520-6106},
journal = {The journal of physical chemistry. B},
}

@article{Taylor_64,
number = {1382},
pages = {383-397},
publisher = {The Royal Society},
title = {Disintegration of Water Drops in an Electric Field},
volume = {280},
year = {1964},
language = {eng},
author = {Geoffrey Taylor},
issn = {1364-5021},
journal = {Proceedings of the Royal Society. A, Mathematical, physical, and engineering sciences},
}

@article{10.1063/1.4821137,
    author = {Crowdy, Darren G.},
    title = {Surfactant-induced stagnant zones in the Jeong-Moffatt free surface Stokes flow problem},
    journal = {Physics of Fluids},
    volume = {25},
    number = {9},
    pages = {092104},
    year = {2013},
    month = {09},
    issn = {1070-6631},
    doi = {10.1063/1.4821137},
    url = {https://doi.org/10.1063/1.4821137},
    eprint = {https://pubs.aip.org/aip/pof/article-pdf/doi/10.1063/1.4821137/16066941/092104_1_online.pdf},
}


\bsp	
\label{lastpage}
\end{document}